%% file: main.tex
\documentclass[lettersize,journal]{IEEEtran}
\usepackage{amsmath,amsfonts}
\usepackage{algorithmic}
\usepackage{algorithm}
\usepackage{array}
\usepackage[caption=false,font=normalsize,labelfont=sf,textfont=sf]{subfig}
\usepackage{textcomp}
\usepackage{stfloats}
\usepackage{url}
\usepackage{verbatim}
\usepackage{graphicx}
\usepackage{cite}
\usepackage{booktabs}
\usepackage{multirow}
\usepackage{fontawesome}
\usepackage{cleveref}
\usepackage{tabularx}
\crefname{section}{§}{§§}
\Crefname{section}{§}{§§}
\begin{document}

\title{Large Language Models for Network Intrusion Detection Systems: Foundations, Implementations, and Future Directions}

\author{Shuo Yang, Xinran Zheng, Xinchen Zhang, Jinfeng~Xu, Jinze Li, \\ Donglin Xie, Weicai Long and Edith~C.H.~Ngai$^{*}$,~\IEEEmembership{Senior Member,~IEEE}

\thanks{S. Yang, X. Zhang, J. Xu, J. Li and E. C.H. Ngai are with Department of Electrical and Electronic Engineering, The University of Hong Kong, Hong Kong ASR, China (shuo.yang@connect.hku.hk).}
\thanks{X. Zheng is with Department of Electronic Engineering, Tsinghua University, Beijing, China. D. Xie is with National Institute of Health Data Science, Peking University, Beijing, China. W. Long is with Data Science and Analytics Thrust, Hong Kong University of Science and Technology (Guangzhou), Guangzhou, China.}
\thanks{$^{*}$ Corresponding author: Edith~C.H.~Ngai (chngai@eee.hku.hk).}
}

\markboth{Preprint}%
{Shell \MakeLowercase{\textit{et al.}}: A Sample Article Using IEEEtran.cls for IEEE Journals}

\maketitle

\begin{abstract}

Large Language Models (LLMs) have revolutionized various fields with their exceptional capabilities in understanding, processing, and generating human-like text. This paper investigates the potential of LLMs in advancing Network Intrusion Detection Systems (NIDS), analyzing current challenges, methodologies, and future opportunities. It begins by establishing a foundational understanding of NIDS and LLMs, exploring the enabling technologies that bridge the gap between intelligent and cognitive systems in AI-driven NIDS. While Intelligent NIDS leverage machine learning and deep learning to detect threats based on learned patterns, they often lack contextual awareness and explainability. In contrast, Cognitive NIDS integrate LLMs to process both structured and unstructured security data, enabling deeper contextual reasoning, explainable decision-making, and automated response for intrusion behaviors. Practical implementations are then detailed, highlighting LLMs as processors, detectors, and explainers within a comprehensive AI-driven NIDS pipeline. Furthermore, the concept of an LLM-centered Controller is proposed, emphasizing its potential to coordinate intrusion detection workflows, optimizing tool collaboration and system performance. Finally, this paper identifies critical challenges and opportunities, aiming to foster innovation in developing reliable, adaptive, and explainable NIDS. By presenting the transformative potential of LLMs, this paper seeks to inspire advancement in next-generation network security systems.

\end{abstract}

\begin{IEEEkeywords}
Large Language Models, Intrusion Detection Systems, Network Security, Generative Artificial Intelligence.
\end{IEEEkeywords}

\input{Sec/1.Intro}
\input{Sec/2.Background}
\input{Sec/3.Pipeline}
\input{Sec/4.Method}
\input{Sec/5.Future}
\input{Sec/6.Conclusion}

% \section*{Acknowledgments}
% This should be a simple paragraph before the References to thank those individuals and institutions who have supported your work on this article.

\bibliographystyle{IEEEtran}
\bibliography{reference}

% \newpage

 

% \end{IEEEbiography}



\vfill

\end{document}

%% file: Sec/1.Intro.tex
\section{Introduction}

The growing complexity and scale of modern network infrastructures have led to an exponential rise in cyber threats. Network Intrusion Detection Systems (NIDS) play a critical role in safeguarding these infrastructures by monitoring and analyzing network traffic for suspicious activities. However, traditional NIDS, which rely on predefined signatures or statistical methods, often struggle to detect sophisticated or novel attacks. The integration of Artificial Intelligence (AI) has significantly enhanced the capabilities of NIDS, enabling more intelligent detection mechanisms. Among these advancements, Large Language Models (LLMs) have emerged as promising tools due to their unparalleled abilities to understand, process, and generate human-like text.

LLMs, such as GPT-4\footnote{GPT-4: \url{https://openai.com/index/gpt-4}}, LLaMA\footnote{LLaMA: \url{https://github.com/meta-llama}}, and DeepSeek\footnote{DeepSeek: \url{https://www.deepseek.com/}}, have demonstrated remarkable success in diverse applications, ranging from natural language processing (NLP) to reasoning and decision-making tasks. Their ability to extract meaningful insights from vast datasets makes them well-suited for the complex and dynamic nature of intrusion detection. By leveraging LLMs, NIDS can transition from intelligent systems to cognitive systems capable of contextual reasoning, offering faster response and enhanced explainability.

Despite these promising capabilities, research and applications of NIDS are still in their early stages, and systematic exploration remains limited. This paper aims to bridge this gap by investigating the transformative role of LLMs in advancing NIDS. Specifically, we establish a foundational understanding of NIDS and LLMs, tracing their evolution and exploring their potential synergies. We then examine how LLMs can be integrated into key stages of the NIDS pipeline, including data, model, and response. Furthermore, we propose LLM-centered Controller, a novel architecture for orchestrating NIDS operations by coordinating various tools and components to enhance efficiency and effectiveness.

However, integrating LLMs into NIDS presents several challenges, including validity, complexity, and privacy. For instance, LLMs may generate false positives due to biases in training data, struggle with inconsistent threat assessments, and even produce misleading outputs that could hinder security responses. Additionally, their high computational demands would impact real-time processing capabilities, while the analysis of sensitive network traffic raises significant privacy risks. This paper identifies these challenges and outlines future research directions, focusing on enhanced multimodal integration, real-time analysis detection, privacy-preserving collaboration, and multi-agent systems. Through a comprehensive review, we aim to catalyze innovation and development in next-generation NIDS, paving the way for more reliable, adaptive, and explainable network security systems.

The rest of this paper is organized as follows. Section~\ref{sec: background} provides a foundational background on NIDS and LLMs. Section~\ref{sec: transition} examines the transition of AI-driven NIDS from intelligent to cognitive systems, focusing on the general pipeline, existing gaps, and enabling technologies. Section~\ref{sec: implementation} discusses the practical implementation of NIDS with LLMs, including LLM-enhanced Processor, LLM-based Detector, LLM-driven Explainer, and LLM-centered Controller. Section~\ref{sec: discussion} outlines open challenges and future directions, and the conclusion is drawn in Section~\ref{sec: conclusion}.

%% file: Sec/2.Background.tex
\section{Background of NIDS and LLMs}
\label{sec: background}

\subsection{NIDS}

\begin{figure}
    \centering
    \includegraphics[width=0.9\linewidth]{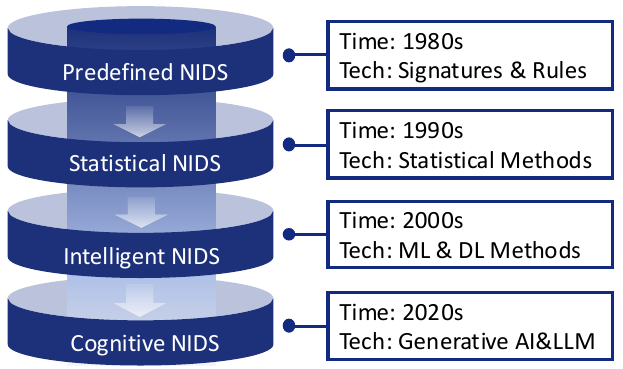}
    \caption{The Roadmap of NIDS.}
    \label{fig: roadmap}
\end{figure}

NIDS are essential to modern cybersecurity, continuously evolving to counter increasingly sophisticated attacks. The roadmap in Fig.~\ref{fig: roadmap} outlines this progression, highlighting key technological advancements.

The earliest \textit{Predefined NIDS} emerged in the late 1980s with systems like Haystack\footnote{Smaha, Stephen E. ``Haystack: An Intrusion Detection System." Fourth Aerospace Computer Security Applications Conference. Vol. 44. 1988.} and SNORT\footnote{https://www.snort.org/}, which relied on static rule-based detection. While effective against known threats, they required frequent updates to address new attack patterns. To overcome this limitation, \textit{Statistical NIDS} appeared in the mid-1990s, introducing anomaly-based detection. Systems such as SPADE\footnote{https://github.com/infosecdr/spade} established statistical baselines of normal network behavior, flagging deviations as potential intrusions. However, high false-positive rates often overwhelm analysts, reducing practicality in dynamic environments.

By the early 2000s, \textit{Intelligent NIDS} leveraged Machine Learning (ML) and Deep Learning (DL) to adapt to evolving threats. Modern tools like Splunk\footnote{https://www.splunk.com/} and Suricata\footnote{https://suricata.io/} moved beyond rule-based detection, using clustering, supervised learning, and ensemble methods to improve accuracy. However, increasing model complexity led to concerns about interpretability. The advent of LLMs has driven the evolution toward \textit{Cognitive NIDS}, capable of processing both structured and unstructured data, including logs, threat intelligence reports, and emails. By leveraging LLMs' reasoning capabilities, Cognitive NIDS enhance contextual awareness, provide actionable security insights, and enable more effective incident response.

\subsection{LLMs}

LLMs represent a major advancement of AI, particularly in NLP domain. These models are trained on vast amounts of data to understand and generate human-like text, excelling in tasks such as summarization, translation, sentiment analysis. Their deep learning architecture, typically based on transformers, enables them to scale to billions of parameters, supporting their impressive performance across various applications.

LLMs have demonstrated broad applicability across industries, highlighting their versatility. In healthcare\footnote{Agent-Hospital: https://www.tairex.cn/agent-hospital}, they assist in analyzing patient records, generating medical reports, aiding preliminary diagnoses, and supporting interactive virtual assistants. In finance\footnote{FinGPT: \url{https://github.com/AI4Finance-Foundation/FinGPT}}, LLMs help analyze complex financial data, generate insights, and improve decision-making processes. In customer service\footnote{ChatGPT-On-CS: \url{https://github.com/cs-lazy-tools/ChatGPT-On-CS}}, they provide 24/7 support, handle inquiries, and enhance user experiences through personalized interactions. These applications illustrate the transformative potential of LLMs across multiple domains, reinforcing their role in driving advancements in cybersecurity and beyond.

%% file: Sec/3.Pipeline.tex
\section{AI-driven NIDS: From Intelligent to Cognitive}
\label{sec: transition}

\begin{figure*}
    \centering
    \includegraphics[width=1\linewidth]{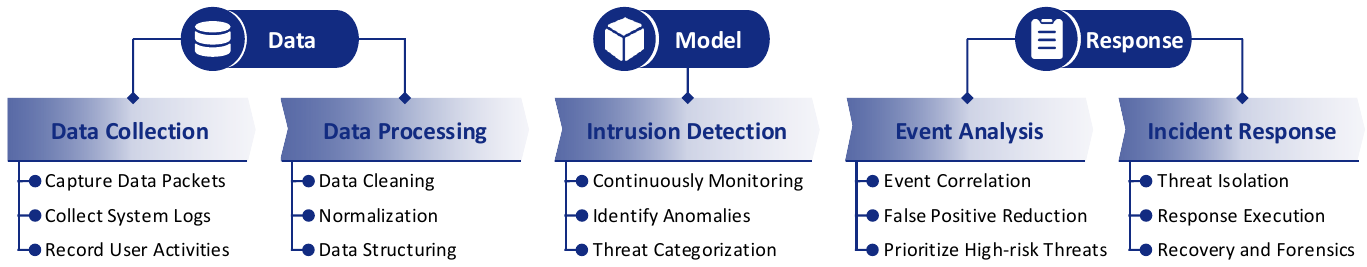}
    \caption{The Pipeline of AI-driven NIDS.}
    \label{fig: pipeline}
\end{figure*}

This section explores the evolution of NIDS from Intelligent to Cognitive, both of which fall under the broader category of AI-driven NIDS. The advancement of AI technologies has significantly enhanced the capabilities of NIDS, enabling them to address increasingly sophisticated security challenges.

\subsection{General Pipeline of AI-driven NIDS}
\label{sec: pipeline}

The effectiveness of AI-driven NIDS lies in their ability to analyze network data, distinguish between normal and anomalous behavior, and respond dynamically to threats. To understand the transition from Intelligent to Cognitive NIDS, we first examine the general pipeline of AI-driven NIDS. Fig.~\ref{fig: pipeline} illustrates five interconnected stages: data collection, data processing, intrusion detection, event analysis, and incident response, forming a backbone of effective AI-driven NIDS.

\subsubsection{Data Collection}

This stage aggregates raw data from packet captures, system logs, application logs, user activity records, and threat intelligence feeds using tools like Wireshark\footnote{Wireshark: https://www.wireshark.org/} and tcpdump\footnote{tcpdump: https://github.com/the-tcpdump-group/tcpdump}. Comprehensive data collection ensures full network visibility, enabling accurate analysis.

\subsubsection{Data Processing}

Raw data is cleansed, normalized, and structured to remove noise and redundancy. This step filters irrelevant traffic, extracts protocol-specific details (e.g., HTTP headers, DNS queries), and enhances data quality, reducing false positives and missed detections.

\subsubsection{Intrusion Detection}

AI-driven NIDS detect both known and emerging threats by continuously monitoring network activities. Intelligent NIDS rely on ML/DL models, while Cognitive NIDS leverage Generative AI and LLMs for deeper contextual understanding, improving detection of complex, multi-stage attacks.

\subsubsection{Event Analysis}

Upon detecting an intrusion, NIDS correlate system logs, user activities, and threat intelligence to assess threat severity and minimize false positives. Integrating external sources like MITRE ATT\&CK\footnote{MITRE ATT\&CK: https://attack.mitre.org/} and Common Vulnerabilities and Exposures (CVE)\footnote{CVE: https://cve.mitre.org/} helps identify zero-day exploits and Advanced Persistent Threat (APTs), improving situational awareness.

\subsubsection{Incident Response}

This stage mitigates threats through automated or manual actions such as isolating compromised systems, blocking malicious traffic, and generating alerts. It also includes forensic investigations and system recovery, ensuring resilience against future attacks.

\subsection{The Gap Between Intelligent and Cognitive NIDS}

\begin{figure*}
    \centering
    \includegraphics[width=1\linewidth]{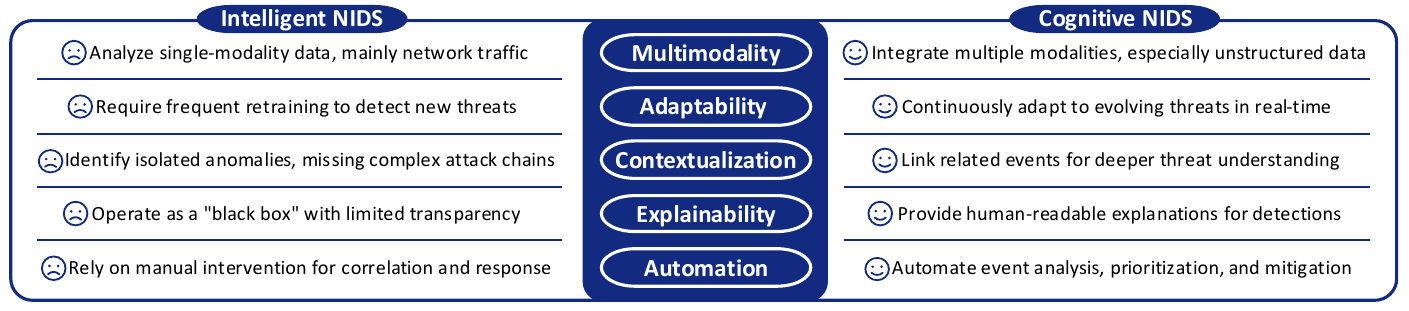}
    \caption{Capability Comparison of Intelligent and Cognitive NIDS.}
    \label{fig: capcom}
\end{figure*}

While Intelligent NIDS utilize ML and DL methods for threat detection, they struggle with unstructured data processing, complex attack detection, explainability, and workflow optimization. Cognitive NIDS, powered by LLMs, overcome these limitations by enhancing contextual understanding, automation, and adaptive decision-making. A comparative overview of their capabilities is summarized in Fig.~\ref{fig: capcom}.

\subsubsection{Unstructured Data Handling}

Intelligent NIDS focus on structured data like IP headers and packet payloads, overlooking insights from unstructured sources such as logs, emails, and threat reports. Cognitive NIDS leverage multimodality, integrating both structured and unstructured data to enhance threat detection. With advanced NLP capabilities, they analyze logs for anomalies, extract Indicators of Compromise (IoCs), and detect phishing or social engineering threats across diverse data formats. This multimodal approach provides a more comprehensive and context-aware view of network security, enabling richer threat intelligence and more accurate intrusion detection.

\subsubsection{Detection of Evolving and Multi-Stage Attacks}

Intelligent NIDS rely on pattern-based detection, often struggling to adapt to evolving threats and multi-stage attacks like APTs, where individual actions appear benign but collectively form a coordinated attack. Cognitive NIDS enhance adaptability by leveraging contextual reasoning and continuously learning from new attack patterns. They dynamically correlate seemingly unrelated events, such as suspicious logins, anomalous file downloads, and data exfiltration, enabling early detection of complex attack chains before significant damage occurs.

\subsubsection{Explainability and Reporting Mechanisms}

Intelligent NIDS, particularly those based on deep learning, often function as ``black boxes", providing little transparency in decision-making. Cognitive NIDS improve explainability by generating natural language descriptions for detected threats, detailing not only what was flagged but also why. They produce human-readable incident reports tailored to different stakeholders, from technical analysts to business leaders, improving decision-making, trust, and response efficiency.

\subsubsection{Automation and Workflow Optimization}

Intelligent NIDS require manual intervention for event correlation, threat prioritization, and response execution, increasing analyst workload and response time. Cognitive NIDS leverage real-time decision-making and contextual automation, dynamically prioritizing threats and even executing mitigation actions, such as isolating compromised endpoints, blocking malicious IPs, and generating automated incident reports. By reducing manual effort, Cognitive NIDS enhance efficiency and accuracy in intrusion detection.

\begin{figure*}
    \centering
    \includegraphics[width=1\linewidth]{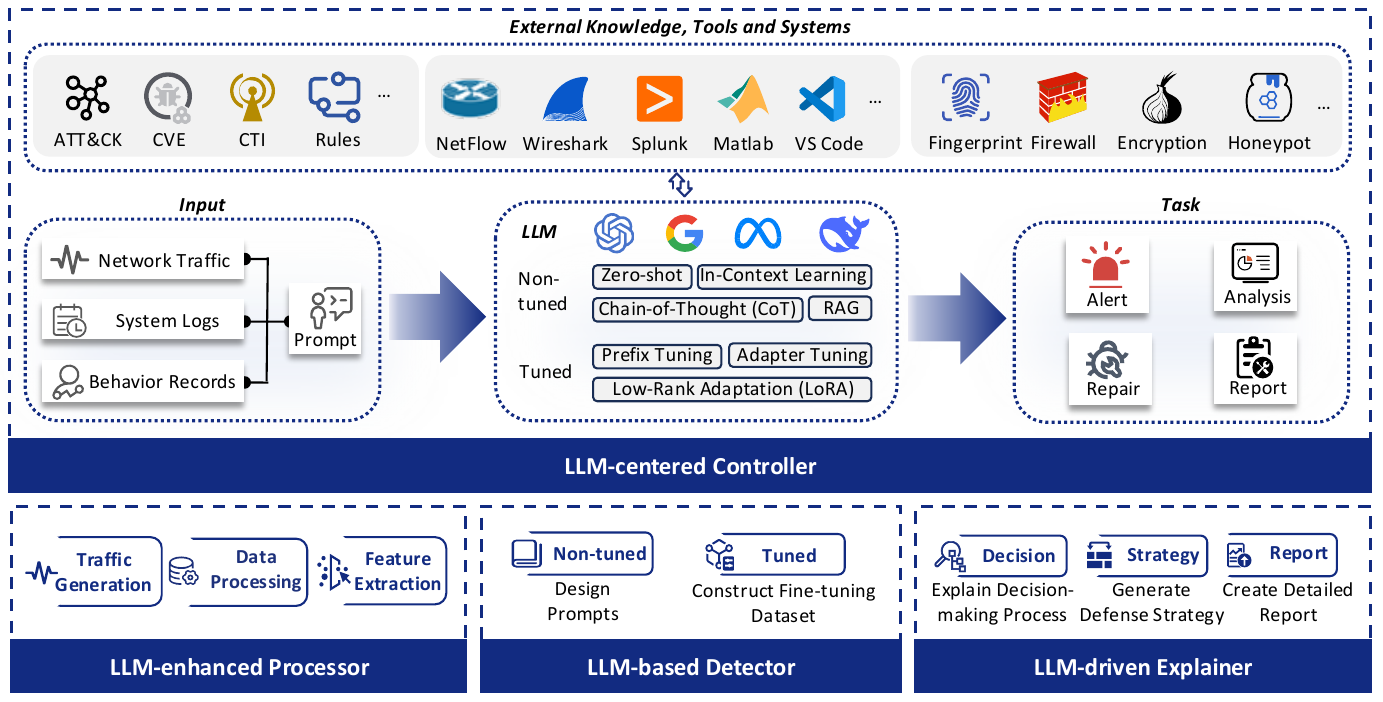}
    \caption{The Overview of Cognitive NIDS.}
    \label{fig: cnids}
\end{figure*}

\subsection{Enabling Techniques for Cognitive NIDS}
\label{sec: tech}

Integrating LLMs to transition from Intelligent to Cognitive NIDS presents unique challenges. Although LLMs have extensive knowledge of common tasks, they may lack domain-specific expertise, and their output can be inherently random. To overcome these challenges, various enabling techniques, categorized into non-tuned and tuned methods, can facilitate the effective integration of LLMs into NIDS.

\subsubsection{Non-tuned Methods}

\paragraph{Zero-shot Prompt} 

Traditional NLP techniques require large labeled datasets for supervised training. In contrast, Zero-shot prompting uses pre-trained LLMs directly for tasks without needing task-specific training. However, this method faces challenges, such as the pre-trained model’s insufficient knowledge of network intrusion detection and biases inherent in the training data.

\paragraph{In-Context Learning (ICL)} 

ICL enhances LLMs by using formatted prompts with task descriptions and examples, allowing the model to learn domain knowledge dynamically without modifying its parameters. This method improves the LLM’s adaptability and task performance by effectively guiding the model in understanding specific requirements and applying domain-specific knowledge.

\paragraph{Chain-of-Thought (CoT)} 

CoT prompts are designed to guide LLMs through complex reasoning tasks by incorporating intermediate reasoning steps. Unlike ICL, which focuses on input-output pairs, CoT prompts help the LLM connect logical steps, enhancing accuracy and stability in solving problems that require abstract or symbolic reasoning.

\paragraph{Retrieval-Augmented Generation (RAG)} 

RAG combines information retrieval with LLMs' generative capabilities, allowing models to retrieve and integrate relevant information from external sources to generate more accurate, context-aware responses. This approach bridges the knowledge gap between general models and up-to-date information, enabling more effective task performance.

\subsubsection{Tuned Methods}

Tuned methods are categorized into Full Fine-Tuning (FFT) and Parameter-Efficient Fine-Tuning (PEFT). FFT updates all model parameters, making it resource-intensive and computationally demanding. In contrast, PEFT modifies only a subset of parameters, providing a more efficient and scalable alternative, particularly suited for network intrusion detection. Below are key PEFT techniques.

\paragraph{Prefix Tuning} 

Prefix tuning introduces task-specific prefix vectors to the input sequence, optimizing the model for specific tasks without altering the original parameters. This approach is efficient, reducing computational overhead and enhancing flexibility and interpretability by adapting the model with minimal changes.

\paragraph{Adapter Tuning} 

Adapter tuning integrates lightweight modules into pre-trained models, enabling task adaptation without altering the model’s core architecture. These modular adapters provide flexibility for experimentation and efficient task optimization.

\paragraph{Low-Rank Adaptation (LoRA)} 

LoRA introduces small, low-rank matrices into critical layers of a model, reducing the number of trainable parameters required for fine-tuning. This method maintains model performance while minimizing storage and computational costs, making it highly suitable for large models, including transformers.

The above techniques could equip LLMs with domain-specific capabilities for NIDS, enabling their seamless integration and advancing the development of Cognitive NIDS.

%% file: Sec/4.Method.tex
\section{The Implementation of NIDS with LLMs}
\label{sec: implementation}

This section examines the role of LLMs in advancing Cognitive NIDS. With their extensive internal knowledge and advanced reasoning capabilities, LLMs can enhance or integrate with existing NIDS to improve efficiency and effectiveness at various stages of the pipeline discussed in \cref{sec: pipeline}. Fig.~\ref{fig: cnids} provides an overview of Cognitive NIDS, which comprises four promising components: \textit{LLM-enhanced Processor} (\cref{sec: llmp}), \textit{LLM-based Detector} (\cref{sec: llmd}), \textit{LLM-driven Explainer} (\cref{sec: llme}), and \textit{LLM-centered Controller} (\cref{sec: llmc}). The following sections provide a comprehensive analysis of these four components, with representative implementations summarized in Table~\ref{tab: llmfornids}.

\begin{itemize}
    \item LLM-enhanced Processor improves data processing efficiency and extracts informative features to enhance threat identification.
    \item LLM-based Detector leverages LLMs to detect complex attack patterns, significantly enhancing intrusion detection performance.
    \item LLM-driven Explainer enhances explainability by interpreting decision-making processes, generating defense strategies, and producing detailed reports.
    \item LLM-centered Controller optimizes workflow orchestration, facilitating seamless collaboration between various system components.
\end{itemize}

By leveraging LLMs, NIDS benefit from enhanced data processing, sophisticated detection capabilities, improved explainability, and optimized workflows. Cognitive NIDS provide stronger, more adaptive protection against the ever-evolving cybersecurity threat landscape.

\input{Tab/summary}

\subsection{LLM-enhanced Processor}
\label{sec: llmp}

The LLM-enhanced Processor leverages LLMs to improve the data processing capabilities of NIDS. In intrusion detection, the monitored data primarily consists of network traffic, user behavior logs, system event logs, application messages, and alerts from security devices. This data serves as the foundation for identifying potential security threats.

\subsubsection{Traffic Generation} 

Network traffic data is fundamental for constructing effective NIDS, with model performance largely dependent on the quality and diversity of the training data. The scarcity of high-quality datasets in network intrusion detection presents significant challenges, especially in real-world applications. Existing datasets are often outdated or insufficient in both quality and quantity, as organizations are reluctant to share data due to privacy concerns or the protection of commercial secrets. As a result, traffic generation has emerged as a valuable approach, utilizing synthetic techniques to simulate network traffic for performance evaluation and benchmarking of NIDS. For instance, Meng~\textit{et al.}~\cite{netgpt} developed NetGPT, which integrates multi-modal traffic modeling by encoding heterogeneous network traffic headers and payloads into a unified text input for traffic understanding and generation tasks. Qu~\textit{et al.}~\cite{trafficgpt} introduced TrafficGPT, which directly generated PCAP files from token lists, achieving high quality and authenticity as demonstrated by metrics like JS divergence. Additionally, Kholgh~\textit{et al.}~\cite{pac-gpt} created PAC-GPT, which fine-tuned GPT-3 to generate ICMP and DNS packets. These developments showcase the feasibility of using generative models such as LLMs to generate high-quality traffic datasets.

\subsubsection{Data Processing}

Transforming raw network data into a usable format often involves complex tasks such as data cleaning, normalization, and integration, particularly when dealing with heterogeneous data from various sources. LLMs, with their advanced semantic understanding and reasoning capabilities, can automate this process by identifying and correcting errors, missing values, and redundant information, thereby enhancing data quality and consistency. For example, Zhang~\textit{et al.}~\cite{lemur} proposed a log parsing framework based on information entropy sampling and CoT prompting, eliminating the need of manual rules and achieving high efficiency in log parsing for downstream tasks. Furthermore, LLMs can identify and correlate data from diverse sources, enabling a more comprehensive understanding of network attacks. Daniel~\textit{et al.}~\cite{nidsattck} demonstrated the use of ChatGPT for labeling NIDS rules with MITRE ATT\&CK techniques, showcasing LLMs' ability to process and integrate multi-source data effectively. In summary, the application of LLMs in data processing significantly reduces the need for manual intervention, enhances data accuracy, and lays a strong foundation for subsequent detection and response strategies.

\subsubsection{Feature Extraction}

LLMs excel at extracting meaningful features from data, which is essential for identifying potential security risks. For instance, Wang~\textit{et al.}~\cite{embedding} used ChatGPT to obtain embeddings from IoT traffic payloads, which were then used to train a deep learning model for IoT traffic anomaly detection. Their results highlighted the importance of the embedding layer in capturing contextual and lexical nuances, which significantly improved detection accuracy and reduced false alarms. Zhang~\textit{et al.}~\cite{zhanghan} applied LLMs for feature selection by ranking the importance of selected features at three levels: very important, kind of important, and not very important. This approach streamlined feature length and minimized the impact of irrelevant features on detection performance. These studies demonstrate that LLMs can significantly enhance feature extraction, improving the detection of anomalous behaviors and boosting the performance of intrusion detection models.

\subsection{LLM-based Detector}
\label{sec: llmd}

While several studies have explored the integration of native LLMs with cybersecurity, only a few have proposed specific implementations for intrusion detection. Building on the techniques discussed in Section~\ref{sec: tech}, we introduce these advancements from both tuned and non-tuned perspectives.

\subsubsection{Non-tuned LLM-based Detector}

~\cite{tune} have applied Zero-shot prompting to directly use LLMs for detection tasks. However, a key challenge is that pre-trained LLMs may lack sufficient knowledge of network attacks to perform complex detection tasks effectively. To address this limitation, Zhang~\textit{et al.}~\cite{zhanghan} employed a simple ICL approach by providing labeled examples to the model using GPT-4. Their method achieved over 90\% accuracy on a simple dataset containing only five attack types, demonstrating that ICL can effectively improve the detection performance of LLMs. In another work, Bui~\textit{et al.}~\cite{pmragft} enhanced frozen LLMs by supplementing them with useful information from ChromaDB\footnote{ChromaDB: https://www.trychroma.com/} and Langchain framework\footnote{Langchain: https://www.langchain.com/}, such as malicious payloads and threat intelligence related to suspected attacker IPs, achieving significant improvements over few-shot prompting. These studies suggest that leveraging the inherent capabilities of LLMs, along with supplemental information, can significantly enhance their performance for intrusion detection without updating the pre-trained model parameters. 

\subsubsection{Tuned LLM-based Detector}

While pre-trained LLMs excel in a wide range of language-understanding tasks, they may not perform optimally on specific tasks such as intrusion detection. To address this, fine-tuning—further training on task-specific data—can be applied, enabling the model to understand and perform detection tasks more effectively. Compared to relying solely on prompting, tuned methods offer a more direct and efficient optimization by adjusting model parameters specifically for the task at hand. Fine-tuned LLMs integrate more seamlessly with existing NIDS, reducing dependence on complex prompts and improving performance. For example, Houssel~\textit{et al.}~\cite{tune} fine-tuned the LLaMA3 model using the NetFlow dataset, applying Odds Ratio Preference Optimization (ORPO) and Kahneman-Tversky Optimization (KTO) techniques. They proposed that LLMs should serve as complementary solutions to state-of-the-art NIDS, enhancing their capabilities. Fine-tuning also allows smaller models to achieve performance on par with larger ones, reducing computational costs and latency during inference. Rigaki~\textit{et al.}~\cite{loratune} used LoRA to fine-tune a 7-billion-parameter pre-trained LLM (Zephyr-7b-$\beta$), achieving performance comparable to that of more powerful models such as GPT-4. These studies show that fine-tuning not only improves model performance for network intrusion detection but also optimizes resource utilization, maintaining or even enhancing performance with lower computational overhead.

\subsection{LLM-driven Explainer}
\label{sec: llme}

The LLM-driven Explainer leverages LLMs to enhance decision-making and strategy formulation in cybersecurity. By analyzing current security incidents and historical data, it provides decision suggestions, response strategies, and automated reports that summarize detected events, actions, and justifications. This improves the explainability and transparency of NIDS. In this section, we explore how LLMs contribute to NIDS explainability across three key dimensions: decision, strategy, and report.

\subsubsection{Decision-level Explainability}

At the decision level, LLMs correlate multi-source data and explain why a specific threat is flagged by detailing the key features, patterns, or anomalies that influence the detection decision. By incorporating both traditional network metrics and semantic information from network communications, LLMs offer a more comprehensive analysis of network behavior. Through in-context learning, LLMs could enhance decision explainability by understanding event relationships and contextual information. Ziems~\textit{et al.}~\cite{feature1} demonstrated that LLM-generated explanations for decision tree-based NIDS models align closely with human assessments in terms of readability, quality, and contextual knowledge, aiding in a clearer understanding of decision boundaries. Additionally, Khediri~\textit{et al.}~\cite{feature2} integrated SHAP explanations with LLM-generated descriptions, providing feature importance and human-readable justifications for NIDS model predictions.

\subsubsection{Strategy-level Explainability}

At the strategy level, LLMs clarify how the system approaches various types of threats and multi-stage attacks. The LLM-driven Explainer generates response strategies such as isolating suspicious users, restricting access, or monitoring traffic. These LLM-generated recommendations enable security teams to take appropriate actions, improving strategy accuracy and response efficiency. Bui~\textit{et al.}~\cite{pmragft} introduced a method to measure the contribution of each token to a predicted output label, allowing for concise incident summaries. This contextual information enables human analysts to interact with the model for further Q\&A and assistance. Jüttner~\textit{et al.}~\cite{chatids} developed ChatIDS, an early-stage security alert explanation system powered by ChatGPT, demonstrating the potential of LLMs in providing intuitive interpretations of security alerts and actionable security measures.

\subsubsection{Report-level Explainability}

In cybersecurity incident response, LLMs can generate detailed analytical reports and visualizations, supporting security teams with comprehensive event descriptions, detection processes, impact assessments, and recommended mitigation measures. Automating report generation significantly reduces the burden of manual documentation while improving situational awareness. By analyzing historical data, LLMs can identify trends and patterns in security events, assisting teams in retrospective analyses and predicting future attack behaviors. This enables organizations to proactively prepare for emerging threats. Ali~\textit{et al.}~\cite{report1} developed HuntGPT, a system that provides a dashboard summarizing attack types, key features influencing model decisions, and their impacts. Additionally, it allows users to interactively obtain real-time response recommendations and detailed investigative reports.

\subsection{LLM-centered Controller}
\label{sec: llmc}

The LLM-centered Controller serves as the orchestrator of the intrusion detection workflow, ensuring seamless collaboration among various tools and components to maximize system efficiency and effectiveness. As illustrated in Fig.~\ref{fig: cnids}, the Controller leverages LLMs to manage processes across all stages, from data processing to incident response. It facilitates communication between modules, dynamically adjusts system configurations, and ensures smooth task execution. For instance, by integrating with traffic collection tools like Wireshark, the LLM can capture and analyze network traffic in real-time, extract key features, and automate traffic processing, optimizing data utilization and analysis. 

Beyond data handling, the Controller enhances incident response by coordinating actions across multiple security tools, such as fingerprint recognition, encryption systems, and firewalls, providing a unified view with visual analytics that enhances interpretability. It can automate the isolation of compromised systems, enforce firewall rules for mitigation, or trigger in-depth forensic investigations on flagged endpoints. This integrated approach improves security posture assessment, reduces manual effort, shortens response times, and strengthens the organization’s overall cybersecurity defenses. Additionally, it dynamically updates threat intelligence by retrieving information from internal and external knowledge bases such as MITRE ATT\&CK, CVE, and Cyber Threat Intelligence (CTI\footnote{CTI: https://github.com/OpenCTI-Platform/opencti}). This continuous adaptation enables the system to respond effectively to evolving security challenges. By serving as an intelligent unifying orchestrator, the LLM-centered Controller ensures that all NIDS components work harmoniously, adapting to dynamic threats and delivering real-time, robust protection against sophisticated attacks.

Early research has demonstrated the feasibility of this architecture. Li~\textit{et al.}~\cite{agent} proposed IDS-AGENT, an LLM-based intrusion detection agent that employs an iterative reasoning-action pipeline. IDS-AGENT extracts traffic data, preprocesses it, calls various machine learning models for classification, retrieves knowledge from internal and external sources, and generates final detection reasoning. While IDS-AGENT achieves competitive performance compared to traditional NIDS, its focus remains on managing detection models rather than orchestrating the entire intrusion detection lifecycle or integrating external knowledge, tools, and systems comprehensively.

%% file: Tab/summary.tex
\begin{table*}[]
\footnotesize
\setlength\tabcolsep{2pt}
\caption{Implementation Summary of NIDS with LLMs}
\label{tab: llmfornids}
\centering
\begin{tabular}{@{}cccc|cc|ccc|ccc@{}}
\toprule
\multirow{3}{*}{Ref.} & \multicolumn{3}{c}{LLM-enhanced Processor} & \multicolumn{2}{c}{LLM-based Detector} & \multicolumn{3}{c}{LLM-driven Explainer} & \multirow{3}{*}{\begin{tabular}[c]{@{}c@{}}LLM-centered \\ Controller\end{tabular}} & \multirow{3}{*}{LLMs} & \multirow{3}{*}{Dataset} \\ 
\cmidrule(lr){2-4} \cmidrule(lr){5-6} \cmidrule(lr){7-9}
& \begin{tabular}[c]{@{}c@{}} Traffic \\ Generation\end{tabular} & \begin{tabular}[c]{@{}c@{}}Data \\ Processing\end{tabular} & \begin{tabular}[c]{@{}c@{}}Feature \\ Extraction\end{tabular} & Non-tuned & Tuned & Decision & Strategy & Report & & & \\\midrule
\cite{netgpt} & \faCircle & \faCircleO & \faCircleO & \faCircleO & \faCircleO & \faCircleO & \faCircleO & \faCircleO & \faCircleO & GPT-2& ISXW2016 etc. \\
\cite{trafficgpt} & \faCircle & \faCircleO & \faCircleO & \faCircleO & \faCircleO & \faCircleO & \faCircleO & \faCircleO & \faCircleO & NetGPT& ISCXTor2016 etc. \\
\cite{pac-gpt} & \faCircle & \faCircleO & \faCircleO & \faCircleO & \faCircleO & \faCircleO & \faCircleO & \faCircleO & \faCircleO & GPT-3& ToN IoT Dataset  \\
\cite{lemur} & \faCircleO & \faCircle & \faCircleO & \faCircleO & \faCircleO & \faCircleO & \faCircleO & \faCircleO & \faCircleO & GPT-4& LogHub \\
\cite{nidsattck} & \faCircleO & \faCircle & \faCircleO & \faCircleO & \faCircleO & \faCircleO & \faCircleO & \faCircleO & \faCircleO & GPT-3.5, GPT-4& N/A \\
\cite{embedding} & \faCircleO & \faCircleO & \faCircle & \faCircleO & \faCircleO & \faCircleO & \faCircleO & \faCircleO & \faCircleO & GPT-4 & Private \\
\cite{zhanghan} & \faCircleO & \faCircleO & \faCircle & \faCircle & \faCircleO & \faCircleO & \faCircleO & \faCircleO & \faCircleO & GPT-4, LLaMA2 etc.& DTL-IDS 5G\\
\cite{pmragft} & \faCircleO & \faCircleO & \faCircleO & \faCircle & \faCircle & \faCircle & \faCircle & \faCircleO & \faCircleO & GPT-4, Mistral etc. & Private \\
\cite{tune} & \faCircleO & \faCircleO & \faCircleO & \faCircle & \faCircle & \faCircle & \faCircleO & \faCircleO & \faCircleO & GPT-4, LLaMA3 & UNSW-NB15 etc. \\
\cite{loratune} & \faCircleO & \faCircleO & \faCircleO & \faCircleO & \faCircle & \faCircleO & \faCircleO & \faCircleO & \faCircleO & Zephyr-7b-$\beta$ & Private  \\
\cite{feature1} & \faCircleO & \faCircleO & \faCircleO & \faCircleO & \faCircleO & \faCircle & \faCircleO & \faCircleO & \faCircleO & GPT-4 & NF-BoT \\
\cite{feature2} & \faCircleO & \faCircleO & \faCircleO & \faCircleO & \faCircleO & \faCircle & \faCircleO & \faCircleO & \faCircleO & Mistral & CICIDS2017 \\
\cite{chatids}& \faCircleO & \faCircleO & \faCircleO & \faCircleO & \faCircleO & \faCircleO & \faCircle & \faCircleO & \faCircleO & GPT-3.5 & N/A  \\
\cite{report1} & \faCircleO & \faCircleO & \faCircleO & \faCircleO & \faCircleO & \faCircle & \faCircle & \faCircle & \faCircleO & GPT-3.5 & KDD99 dataset  \\
\cite{agent}& \faCircleO & \faCircle & \faCircleO & \faCircle & \faCircleO & \faCircle & \faCircleO & \faCircleO & \faCircle & GPT-3.5, GPT-4o etc. & ACI-IoT23 etc. \\ \bottomrule
\end{tabular}
\end{table*}

%% file: Sec/5.Future.tex
\section{Challenges and Future Directions}
\label{sec: discussion}
This section discusses the challenges of integrating LLMs into NIDS and outlines key future directions to maximize their potential in intrusion detection.

\subsection{Challenges}

\subsubsection{Validity}

A primary challenge in utilizing LLMs for NIDS is the reliability and consistency of their outputs. Due to the inherent unpredictability of LLMs, threat assessments may be compromised by hallucinations\footnote{Huang, Lei, et al. ``A Survey on Hallucination in Large Language Models: Principles, Taxonomy, Challenges, and Open Questions." ACM Transactions on Information Systems 43.2 (2025): 1-55.}, where models generate inaccurate, misleading, or contextually irrelevant outputs. Additionally, biased training datasets can skew detection results, leading to false positives or blind spots in identifying specific attack patterns. Another critical issue is the security of LLMs themselves—adversarial attacks could manipulate or deceive the model, allowing malicious actors to evade detection.

\subsubsection{Complexity}

Deploying LLMs in real-time network environments poses significant computational and infrastructural challenges. Large-scale models require substantial processing power, memory, and storage, limiting their feasibility in resource-constrained environments such as edge networks and IoT ecosystems. Moreover, high latency in processing vast amounts of network traffic would delay detection and response time, potentially leaving systems vulnerable to fast-moving cyber threats.

\subsubsection{Privacy}

Integrating LLMs into NIDS raises privacy concerns, particularly regarding the handling of sensitive data within network traffic and logs. Since LLMs process extensive datasets, they may inadvertently expose private or confidential information, violating data protection regulations such as GDPR\footnote{GDPR: https://gdpr-info.eu/}. Additionally, security risks associated with data leakage during training and inference must be addressed to prevent unauthorized access or misuse of sensitive information.

\subsection{Future Directions}

\subsubsection{Multimodal Integration for Robust Threat Detection}

One promising avenue for future research is the integration of multimodal data to enhance the performance and reliability of NIDS with LLMs. By combining multiple data sources, such as network traffic, system logs, behavioral analytics, and external threat intelligence, LLMs can provide a more contextualized and accurate understanding of threats. This integration enhances detection accuracy, reduces false positives, and ensures models are not overly dependent on single-modality inputs, mitigating biases and improving generalization.

\subsubsection{Optimizing Real-Time Intrusion Detection for Edge Networks}

For LLMs to be effectively applied in edge networks, advancements in real-time analysis and detection are crucial. Future research should focus on developing lightweight models (e.g. Small Language Model\footnote{Zhang, Peiyuan, et al. ``TinyLLaMA: An Open-Source Small Language Model." arXiv preprint arXiv:2401.02385 (2024).}) that require fewer computational resources and while maintaining high performance. Techniques such as model quantization, knowledge distillation, and efficient architecture can help reduce latency in threat detection, making real-time systems more feasible. Deploying Small Language Models at the edge (e.g. IoT gateways, routers, and mobile devices) will enable low-latency detection, improving response times and making Cognitive NIDS feasible in resource-constrained environments.

\subsubsection{Secure and Privacy-Preserving Threat Intelligence Collaboration}

To balance privacy and detection effectiveness, the NIDS with LLMs should integrate privacy-preserving techniques such as federated learning, homomorphic encryption, and secure multi-party computation. These approaches allow multiple organizations to collaborate on improving intrusion detection models without sharing sensitive data. Additionally, Zero-Trust Architectures (ZTA\footnote{ZTA: https://www.nist.gov/publications/zero-trust-architecture}) and confidential computing can ensure that LLM inference is performed securely, reducing risks associated with data exposure during processing. By integrating LLMs with existing security tools, such as firewalls, SIEMs, and endpoint detection systems, Cognitive NIDS can provide more effective threat intelligence sharing and adaptive security measures while maintaining privacy compliance.

\subsubsection{Multi-Agent Systems for Collaborative Intrusion Detection}
A promising direction for Cognitive NIDS is the development of multi-agent systems (MAS\footnote{Talebirad, Yashar, and Amirhossein Nadiri. "Multi-Agent Collaboration: Harnessing the Power of Intelligent LLM Agents." arXiv preprint arXiv:2306.03314 (2023).}), where autonomous LLM-agents collaborate to enhance threat detection, response, and decision-making. Unlike centralized models, MAS distribute security tasks among specialized agents, each analyzing different data sources such as network traffic, logs, and behavioral patterns. These agents share real-time intelligence, improving adaptive threat detection and cross-context correlation. By operating in parallel, MAS reduces computational overhead, enhances scalability, and enables faster incident response. Additionally, coordinated agents can autonomously mitigate threats, such as isolating compromised nodes or adjusting firewall rules. Future research should focus on optimizing MAS architectures to achieve greater accuracy, automation, and resilience against evolving cyber threats. 

%% file: Sec/6.Conclusion.tex
\section{Conclusion}
\label{sec: conclusion}

This paper has examined the transformative role of Large Language Models (LLMs) in advancing Network Intrusion Detection Systems (NIDS), highlighting their potential to transition the field from intelligent to cognitive systems. By exploring practical implementations, we showcase LLMs as processors, detectors, explainers, and controllers, with the LLM-centered Controller emphasizing the orchestration capabilities to streamline workflows and enhance system efficiency. While challenges such as validity, complexity, and privacy remain, proposed future research directions—including multimodal integration, real-time analysis, privacy-preserving collaboration, and multi-agent systems—provide a pathway to address these issues. We hope this work inspires further efforts to harness LLMs for reliable, adaptive, and explainable NIDS.